# Current-induced non-uniform enhancement of sheet resistance in Ar+ irradiated $SrTiO_3$

Debangsu Roy[1], Yiftach Frenkel[1], Sagi Davidovitch[1], Eylon Persky[1], Noam Haham[1], Marc Gabay[2], Beena Kalisky[1], and Lior Klein[1]

*[1]Department of Physics, Nano-magnetism Research Center, Institute of Nanotechnology and Advanced Materials, Bar-Ilan University,Ramat-Gan 52900, Israel.*

*[2] Laboratoire de Physique des Solides, Universite Paris-Sud and CNRS, Batiment 510, 91450 Osray, France.*

The sheet resistance $R_s$ of $Ar^+$ irradiated $SrTiO_3$ in patterns with a length scale of several microns increases significantly below ~40 K in connection with driving currents exceeding a certain threshold. The initial lower $R_s$ is recovered upon warming with accelerated recovery around 70 and 160 K. Scanning SQUID microscopy shows local irreversible changes in the spatial distribution of the current with a length scale of several microns. We attribute the observed non-uniform enhancement of $R_s$ to the attraction of the charged single- and di- oxygen vacancies by the crystallographic domain boundaries in $SrTiO_3$. The boundaries which are nearly ferroelectric below 40 K are polarized by the local electrical field associated with the driven current and the clustered vacancies which suppress conductivity in their vicinity yield a noticeable enhancement in the device resistance when the current path width is on the order of the boundary extension. The temperatures of accelerated conductivity recovery are associated with the energy barriers for the diffusion of the two types of vacancies.

# I. INTRODUCTION

Perovskites attract tremendous interest for their wide range of intriguing properties including magnetism, high $T_c$ superconductivity and multiferroicity [1]. Special attention is paid to heterostructures consisting of $SrTiO_3$ (STO) which is expected to play a central role in oxide based electronics. Of particular interest are quasi- 2D conducting states emerging at the interface between $SrTiO_3$ and different transition metal oxides [2-5] including $LaAlO_3/SrTiO_3$ (LAO/STO). As the emerging phenomena in these heterostructures are commonly attributed to the modified $SrTiO_3$ states, it is important to explore properties of conducting layers in STO surfaces modified by irradiation or gating. [6-8]

The stoichiometric bulk STO is a nonpolar cubic band insulator with a band gap of ~ 3.2 eV. However, upon different treatments high mobility conductivity may be obtained [8]. A common way to make STO conducting is through electron doping realized by introducing oxygen vacancies. When such vacancies form through high temperature annealing in vacuum [9-12] bulk conductivity is obtained. On the other hand, when the oxygen vacancies are introduced by $Ar^+$ irradiation [8] a quasi-2D conducting layer forms close to the surface. Similarly, bulk or surface doping of STO with Nb yields bulk or quasi-2D conductivity, respectively [13,14] . Another route to conductivity is electronic reconstruction at the interface between several layers of $LaAlO_3$ [1] deposited on STO which leads to the formation of a two-dimensional conducting layer. In addition, STO can be doped electrostatically [15,16] using electric double layer transistor (EDLT) [17] configuration employing a polymer electrolyte [10] thus making STO conducting.

Here we report current-induced sheet resistance ($R_s$) increase in patterned STO samples with a length scale of several microns. This phenomenon occurs below ~40 K in connection with driving an electrical current through the sample exceeding a certain threshold. The initial as-cooled $R_s$ recovers upon warming, particularly around 70 and 160 K, similar to the behavior exhibited by LAO/STO system [18,19] . To probe the microscopic manifestation of the effect we have used a scanning superconducting quantum interference device (SQUID) and observed spatial changes in magnetic flux pattern prior to and after the $R_s$ increase. The emerging scenario attributes the effect to two factors: crystallographic domain boundaries in STO which exhibit ferroelectric instability below ~40 K and charged single- and di- oxygen vacancies which become mobile at around 70 and 160 K, respectively. Below ~40 K the local electric field associated with current exceeding a certain threshold is capable of polarizing the crystallographic domain boundaries in STO. The polarized boundaries attract the charged oxygen vacancies and cause their clustering. For current path widths on the order of several microns which is similar to the extension of the boundaries such a clustering yields a noticeable decrease in conductivity, as such vacancies locally suppress electrical conductivity. Upon warming, the boundaries lose their polarization at ~40 K; however, the clusters of the oxygen vacancies disperse only at higher temperatures at which the vacancies become mobile. We believe that the emerging scenario is relevant to the understanding of the physics of other quasi 2D electron gas formed on STO substrate which makes our results particularly important for future development of novel STO-based devices.

## II. EXPERIMENTAL DETAILS

Our samples are commercially available two sided polished STO crystals. The substrates are selectively irradiated to define conducting areas in patterns that allow resistivity measurements with current path width that vary between 2 to 20 μm (see a typical pattern geometry in Figure 1). The selective irradiation is obtained by covering the entire sample with a resist and removing it in specific places by using e-beam lithography. The e-beam resist used for patterning has a thickness of 1.5 μm. The thickness of the e-beam resist was chosen such that it survives the irradiation due to $Ar^+$ beam. For irradiation we use an $Ar^+$ beam with a fluence of $10^{15}$ ions /s $cm^2$ accelerated by 4 KeV for 100 seconds. The estimated penetration depth "L" in Angstrom was calculated using the empirical formula [8] $L = 1.1 \times E^{2/3} W / [\rho (Z_i^{1/4} + Z_t^{1/4})^2]$, where E is the energy in eV, W is the atomic weight of the target in amu, ρ is the target density, and $Z_i$ and $Z_t$ are the atomic numbers of the ions and the targets respectively. The calculated penetration depth was found to be ~ 120Å.

The reported effect was studied using transport measurements carried out in Quantum Design PPMS-9. In addition, we have used a scanning SQUID probe to explore the microscopic manifestation of the effect.

## III. RESULTS

Figure 1 presents a typical current-induced increase of $R_s$ at 5 K in irradiated STO. Data are shown for patterns with channel width of 2, 5, and 8 μm. The initial $R_s$ of the patterns is measured with a small probing current of 0.04 μA. Subsequently, the driving current is increased in steps and $R_s$ is measured for different current values. After each current step, $R_s$ is measured with a probing current value of 0.04 μA to obtain the sheet resistance while excluding Joule heating effects (The "oscillations" in $R_s$ seen in Figure 1(d) are related to joule heating when a high current is driven through the sample). Whereas a spontaneous small increase in $R_s$ is occasionally observed at low driving currents and also when probing current is used. A larger increase is observed consistently above a certain current threshold of ~ 0.3 μA without a clear dependence on the channel width, as might be expected. On the other hand, the magnitude of the effect (the current induced increase of $R_s$) strongly decreases with the increase of the channel width. In order to quantify the relative increase in $R_s$ due to driven current we have calculated $(R_s^{\text{after trapping}} - R_s^{\text{before trapping}})/R_s^{\text{before trapping}}$ in percentage. Figure 1(e) depicts the variation of $R_s^{\text{after trap}} - R_s^{\text{before trap}})/R_s^{\text{before trap}}$ with the sample channel width at 5K. It is found that this parameter strongly depends on the channel width and it becomes negligible when the channel width becomes more than ~ 10 microns.

The initial lower $R_s$ is recovered upon warming. Figure 2(a) shows $R_s$ during cooling from room temperature down to 5 K and during warming after current-induced $R_s$ increase at 5 K in a pattern with a channel width of 5 μm. The inset of Figure 2(a) exhibits the increase in $R_s$ due to trapping. We observe that the conductivity recovery is accelerated around 70 and 160 K as depicted in Figure 2(b). In order to understand the reproducibility of the accelerated recovery temperature, a graph in the inset of Figure 2(b) depicts the similar effect as presented in Figure 2(b) for the two others samples. This denote the reproducibility of the transition occurring at 70K and 160K. *The temperature dependence of the effect is demonstrated in Figure 3. In this figure, the sample is cooled with different applied currents. When the applied current is larger than the threshold current required for resistivity increase, the upturn in the resistance can be seen. We note a clear split between the curves below a threshold temperature of ~40 K and it grows continuously as the temperature is decreased.*

To explore the microscopic manifestation of the effect, we have used a scanning SQUID to probe the current flow distribution in a pattern with a channel width of 20 μm before and after a current-induced increase of $R_s$. During measurement an AC bias current (821 Hz) with a varying magnitude was driven through the sample. The generated magnetic flux was measured by the SQUID at different positions in the sample through a 1.8 μm pick up loop using lock in techniques. *All the SQUID measurements presented in this manuscript are performed at 4.2K.* The resultant SQUID image is a convolution of the z component of the magnetic field and the SQUID point spread function. The SQUID response was found to be linear with the applied current and independent of frequency. The choice of the pattern with channel width of 20 μm is based on the size of the pick-up loop and the spatial resolution of our system. Prior to the measurement of scanning SQUID in this pattern, we have carried out the current-induced increase of $R_s$ at 5 K. Current of magnitude 500 μA is found to induce charge trapping in this pattern.

Figure 4(a) shows the spatial distribution of the magnetic flux resulting from a current below the threshold value flowing in the pattern after the initial cool-down. The flux structure shows an inhomogeneous flow of current. To understand the influence of the microscopic changes on to the charge trapping in STO, high resolution SQUID response was collected over a large area of the sample for different magnitudes of the driven current starting from 0.1 mA up to 4 mA. Figure 4(b) and (c) denote the spatial distribution of magnetic flux for different current magnitudes at two separate locations on the studied pattern. It is evident from these figures that spatial changes in the magnetic flux start to appear as shown in Figure 4(b) as soon as the driven current surpasses the threshold value.

Figure 4(c) exhibits no change in the modulation of the flux densities indicating that the phenomena could be related to the microscopic local domain configuration. In other areas of the sample we find traces for a modulation of current over the underlying domain structure, buried under the overall inhomogeneous current flow. In order to better visualize the spatial variation over which the changes are taking places, two-dimensional current density is depicted in Figure 4(d) –(e) (for the driven current starting from 0.1 mA up to 4 mA). To explore the changes in current distribution due to the studied effect, we use its relation to the measured magnetic flux distribution via Biot-Savart law assuming a two dimensional current flow [20]. Considering the Hall bar structure and the quasi 2D conductivity, we approximated the spatial distribution of the current density according to the method described by Roth et al [20]. The modulation of the current distribution is *reconstructed f*rom ac flux image by employing Fourier transform and considering the geometry of the pickup coil. This is extracted from the flux image presented in Figure 4(b) and (c). The region over which the modulation in the current density (for a driven current) is observed, is estimated to be ~4-5 μm. The figure 4(f) shows high resolution images of a selected region before applying a current of 0.5 mA. The flux structure reflects underlying tetragonal domain structure masked by the inhomogeneities in the current flow. All images and profiles are normalized by the applied current.

## IV. DISCUSSION

In this study we show current-induced increase of $R_s$ in irradiated STO. The effect becomes most notable in patterns with a length scale below ~ 10 microns and it occurs below a threshold temperature of ~ 40 K for large enough applied currents. The initial $R_s$ is recovered upon warming with notable accelerated recovery around 70 and 160 K. Scanning SQUID imaging reveals that the $R_s$ enhancement is nonuniform and correlated with local changes on the order of several microns in the current distribution (see Figure 4). The current density changes (for a particular driven current through the sample) by a factor of two within a 20 μm channel and the region over which the modulation in the current density (for a particular driven current) is observed is estimated to be ~4-5 μm. The effect resembles current-induced increase of $R_s$ is LAO/STO [18,19] heterostructure, where it is found that the accelerated recovery near 70 K and 160 K is associated with two well defined energy barriers $Eb_1 = 0.224 \pm 0.003$ eV and $Eb_2 = 0.44 \pm 0.015$ eV [19]. Thus the observed phenomena in STO seems to be closely related to the charge trapping phenomena observed in LAO/STO. A picture emerges from the experimental data reported in this paper which plausibly accounts for the above facts. It includes charged oxygen vacancies of two types and tetragonal domain walls which are shown to affect electrical transport [21]. The upper bound of the domain wall width was estimated to be ~5 μm at 80 K [22].

Single- and di- oxygen vacancies commonly occur in STO [23,24]. Their formation energies are lowest on the surface since less bonds are being broken when a defect is created there. Based on experiments and DFT calculations it was concluded that these defects reside preferentially at dislocations and twins [25-27]. The latter, which have been imaged before [21,28, 29] , correspond to the boundaries between a-, b- and c- oriented crystallographic domains that form below the ferroelastic transition from cubic to tetragonal symmetry at ~105 K, driven by minute rotation of the $TiO_6$ octahedra [30]. It has been reported in the literature that these tetragonal twin domains strongly affect the microscopic electronic properties [21,31]. In the temperature range when vacancies are mobile, they flow along nanoscale filamentary channels and motion appears correlated within bundles of filaments of typical size 1 - 10 μm [25].

At low temperature, (single-, di-) vacancies cannot overcome the migration energy barrier and they form static clusters within the filament bundles. The values of these barriers at the surface, namely 0.14 -0.2 eV and 0.4 eV for single- and di- vacancy motion respectively [24,32,33], closely match the trapping energies that were determined by Seri et al. [19] for LAO/STO and were shown to correspond to two recovery temperatures at 70 K and 160 K as is also observed in the present study. Since vacancy motions are correlated within a bundle, one expects only small variations in the value of the trapping current for sample widths less than 10 μm. Ab initio calculations show that one of the two electrons released by an oxygen vacancy remains in the vicinity of the defect while the other becomes delocalized and participates in the two dimensional electron liquid layer detected near the surface of STO.

A recent report by Ma et al., [28] furthermore reveals that beyond a certain threshold of electric field, ferroelectricity can be induced within the twin walls in STO. In the literature, the simultaneous occurrence of the twin domain wall mobility and of the local instabilities revealing the closeness to a ferroelectric transition is evidenced experimentally at temperature lower than 40 K [30,31].

We suggest that driving a current exceeding a certain threshold below 40 K polarizes the domain walls which increases the density of the charged oxygen vacancies in their vicinity. The electric field generated by the positive charge of the vacancy leads to a spatial variation of the ferroelectric polarization in the wall. This variation entails a larger value of the polar energy than that produced by a uniform polarization. In order to lower the energy, electrons from the 2DEG bind to the vacancies so as to screen their positive charge and cancel their electric field. This causes an increase in Rs. When oxygen vacancies and di-vacancies become mobile upon warming, i.e. above 70K and 160K, the initial higher conductivity is recovered. A schematic description of this scenario is presented in Figure 5.

In conclusion, we identify a low temperature Rs increase in connection with driving an electrical current higher than a certain threshold through the pattern. This effect is predominant below ~40 K and an accelerated recovery near 70 K and 160 K is observed while warming the sample. Microscopic SQUID imaging of the sample at lower temperature before and after the charge trapping reveals the change in the magnetic flux structure. The modulation of the current density prior to and after charge trapping reveals that the microscopic changes are taking place in the region of ~4-5 μm. We understand the observed phenomena by considering the tendency for domain wall polarization below 40 K, the interaction between charged oxygen vacancies and the domain walls, the internal gating effects, and the energy barriers associated with the mobility of the two types of vacancies. Understanding the interplay between domain walls and oxygen vacancies in this system is crucial for understanding electrical transport properties of this system and for designing future STO-based devices.

L.K. acknowledges support by the U.S.-Israel Binational Science Foundation (2012169) and by the Israel Science Foundation founded by the Israel Academy of Sciences and Humanities (533/15). Y.F., E.P., N.H. and B.K. were supported by the European Research Council Grant No. ERC-2014-STG-639792, and Israel Science Foundation Grant No. ISF-1102/13.

References:

[1] J. Chakhalian, J. W. Freeland, A. J. Millis, C. Panagopoulos, & J. M. Rondinelli, Rev. Mod. Phys. **86**, 1189 (2014).

[2] A. Ohtomo, & H. Y. Hwang, Nature (London) **427**, 423 (2004).

[3] Y. Hotta, T. Susaki, & H. Y. Hwang, Phys. Rev. Lett. **99**, 236805 (2007).

[4] Y. Chen, N. Pryds, J. E. Kleibeuker, G. Koster, J. Sun, E. Stamate, B. Shen, G. Rijnders, & S. Linderoth, Nano Lett. **11**, 3774 (2011).

[5] Y. Z. Chen, N. Bovet, F. Trier, D. V. Christensen, F. M. Qu, N. H. Andersen, T. Kasama, W. Zhang, R. Giraud, J. Dufouleur, T. S. Jespersen, J. R. Sun, A. Smith, J. Nygard, L. Lu, B. Büchner, B. G. Shen, S. Linderoth & N. Pryds, Nat. Commun. 4, 1371 (2013).

[6] A. F. Santander-Syro, O. Copie, T. Kondo, F. Fortuna, S. Pailhès, R. Weht, X. G. Qiu, F. Bertran, A. Nicolaou, A. Taleb-Ibrahimi, P. Le Fèvre, G. Herranz, M. Bibes, N. Reyren, Y. Apertet, P. Lecoeur, A. Barthélémy & M. J. Rozenberg, Nature (London) **469**, 189 (2011).

[7] W. Meevasana, P. D. C. King, R. H. He, S-K. Mo, M. Hashimoto, A. Tamai, P. Songsiriritthigul, F. Baumberger & Z-X. Shen, Nat. Mater. 10, 114 (2011).

[8] M. Schultz, & L. Klein, Appl. Phys. Lett. **91**, 151104 (2007).

[9] O. N. Tufte, & P. W. Chapman, Phys. Rev. **155**, 796 (1967).

[10] H. P. R. Frederikse, & W. R. Hosler, Phys. Rev. **161**, 822 (1967).

[11] K. Szot, W. Speier, R. Carius, U. Zastrow, & W. Beyer, Phys. Rev. Lett. **88**, 075508 (2002).

[12] V. E. Henrich, G. Dresselhaus, & H. J. Zeiger, Phys. Rev. B **17**, 4908 (1978).

[13] G. Binnig, , A. Baratoff, , H. E. Hoenig, & J. G. Bednorz, Phys. Rev. Lett. **45**, 1352 (1980).

[14] A. Spinelli, M. A. Torija, C. Liu, C. Jan, and C. Leighton, Phys. Rev. B **81**, 155110 (2010).

[15] H. Nakamura, **&** H. Takagi, Appl. Phys. Lett. **89**, 133504 (2006).

[16] K. Ueno, S. Nakamura, H. Shimotani, A. Ohtomo, N. Kimura, T. Nojima, H. Aoki, Y. Iwasa & M. Kawasaki, Nat. Mater. **7**, 855 (2008).

[17] Y. Lee, C. Clement, J. Hellerstedt, J. Kinney, L. Kinnischtzke, X. Leng, S. D. Snyder, and A. M. Goldman, Phys. Rev. Lett. **106**, 136809 (2011).

[18] S. Seri, M. Schultz, & L. Klein, Phys. Rev. B **86**, 085118 (2012).

[19] S. Seri, M. Schultz, & L. Klein, Phys. Rev. B **87**, 125110 (2013).

[20] B. J. Roth, N. G. Sepulveda, & J. P. Wikswo Jr, J. Appl. Phys. **65**, 361 (1989).

[21] B. Kalisky, E. M. Spanton, H. Noad, J. R. Kirtley, K. C. Nowack, C. Bell, H. K. Sato, Masayuki Hosoda, Y. Xie, Y. Hikita, C. Woltmann, G. Pfanzelt, R. Jany, C. Richter, H. Y. Hwang, J. Mannhart & K. A. Moler, Nat. Mater. 12, 1091 (2013).

[22] T. A. Merz, H. Noad, R. Xu, H. Inoue, W. Liu, Y. Hikita, A. Vailionis, K. A. Moler, & H. Y. Hwang, Appl. Phys. Lett. **108**, 182901 (2016).

[23] D. D. Cuong, B. Lee, K. M. Choi, H.-S. Ahn, S. Han, and J. Lee, Phys. Rev. Lett. **98**, 115503 (2007).

[24] S. J. Matthew, Marshall, A. E. Becerra-Toledo, L. D. Marks, & M. R. Castell, Defects at Oxide Surfaces Series: Springer Series in Surface Sciences., **58** Chapter 11 (2015).

[25] K. Szot, W. Speier, G. Bihlmayer, & R. Waser, Nat. Mater. **5**, 312 (2006).

[26] D. Marrocchelli, L. Sun , & B. Yildiz, J. Am. Chem. Soc. **137**, 4735 (2015).

[27] S. P. Waldow, & R. A. D. Souza, ACS Appl. Mater. Interfaces **8**, 12246 (2016).

[28] H. J. H. Ma, S. Scharinger, S. W. Zeng, D. Kohlberger, M. Lange, A. Stöhr, X. R. Wang, T. Venkatesan, R. Kleiner, J. F. Scott, J. M. D. Coey, D. Koelle, & Ariando, Phys. Rev. Lett. **116**, 257601 (2016).

[29] M. Honig, J. A. Sulpizio, J. Drori, A. Joshua, E. Zeldov & S. Ilani, Nat. Mater. 12, 1112 (2013).

[30] H. Unoki, & T. Sakudo, J. Phys. Soc. Jpn. **23**, 546 (1967).

[31] J. F. Scott, E. K. H. Salje, & M. A. Carpenter, Phys. Rev. Lett. **109**, 187601 (2012).

[32] A. V. Bandura, R. A. Evarestov, & D. D. Kuruch, Integr Ferroelectr **123**, 1 (2010).

[33] Y. F. Zhukovskii , E. A. Kotomin , S. Piskunov , Y. A. Mastrikov & D. E. Ellis, Ferroelectrics. **379**, 191 (2009).

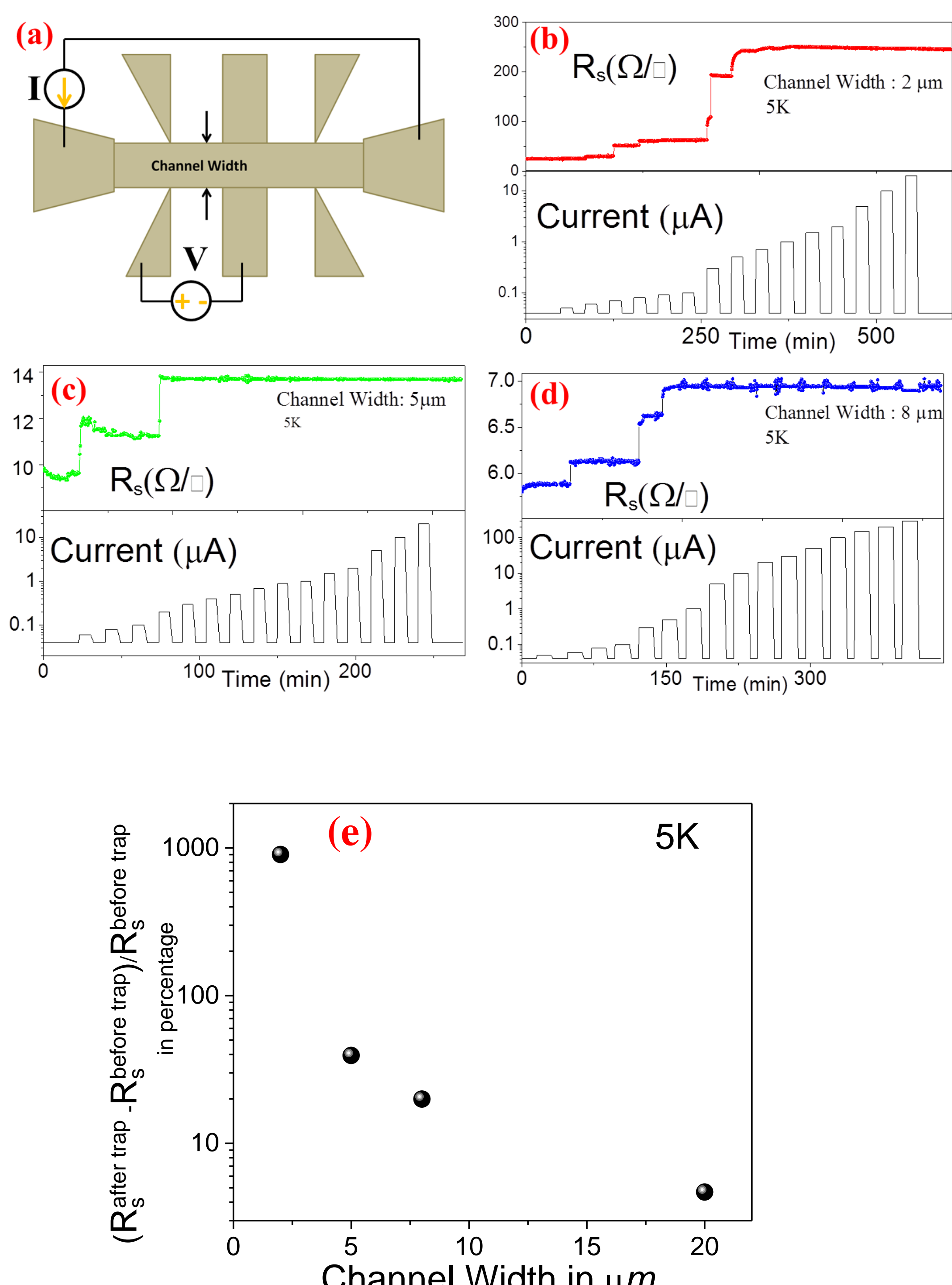

FIG. 1: (a) A sketch of a typical pattern. $R_s$ as a function of time at 5 K as the current is increased in steps for patterns with channel width of 2 μm (b), 5 μm (c) and 8 μm (d). Between current steps, $R_s$ is also measured with a probe current of 0.04 μA. (e) Variation of $(R_s^{after\ trap} - R_s^{before\ trap})/R_s^{before\ trap}$ with the channel width.

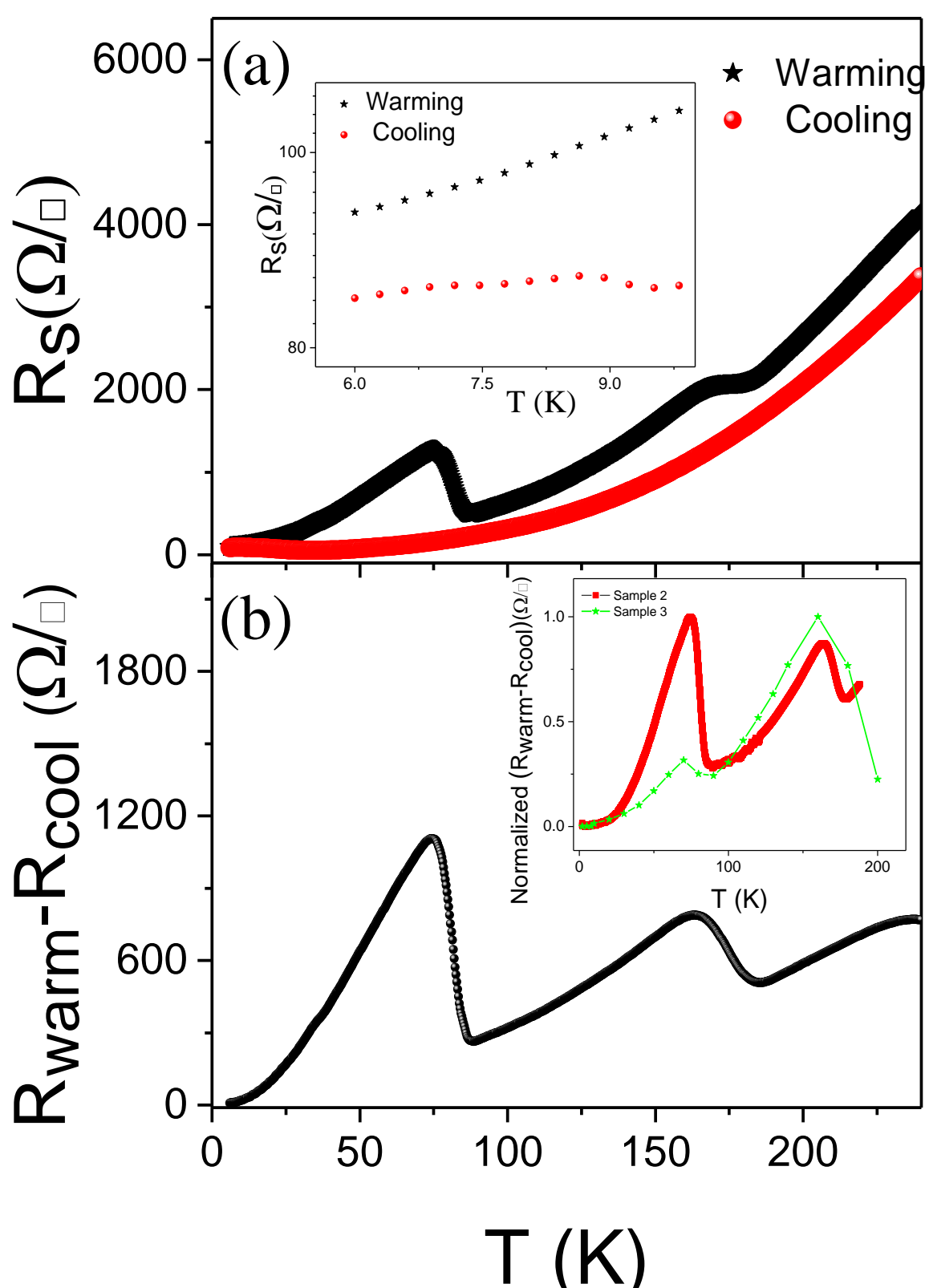

FIG. 2: Temperature dependence of $R_s$ during cooling (red legend) and during warming (black legned) after $R_s$ was increased at a low temperature. Inset depict the increase in $R_s$ at low temperature due to charge trapping. (b) Temperature dependence of R warmning - R cooling. Inset shows the reproducibility of the accelerated recovery temperature for two different samples.

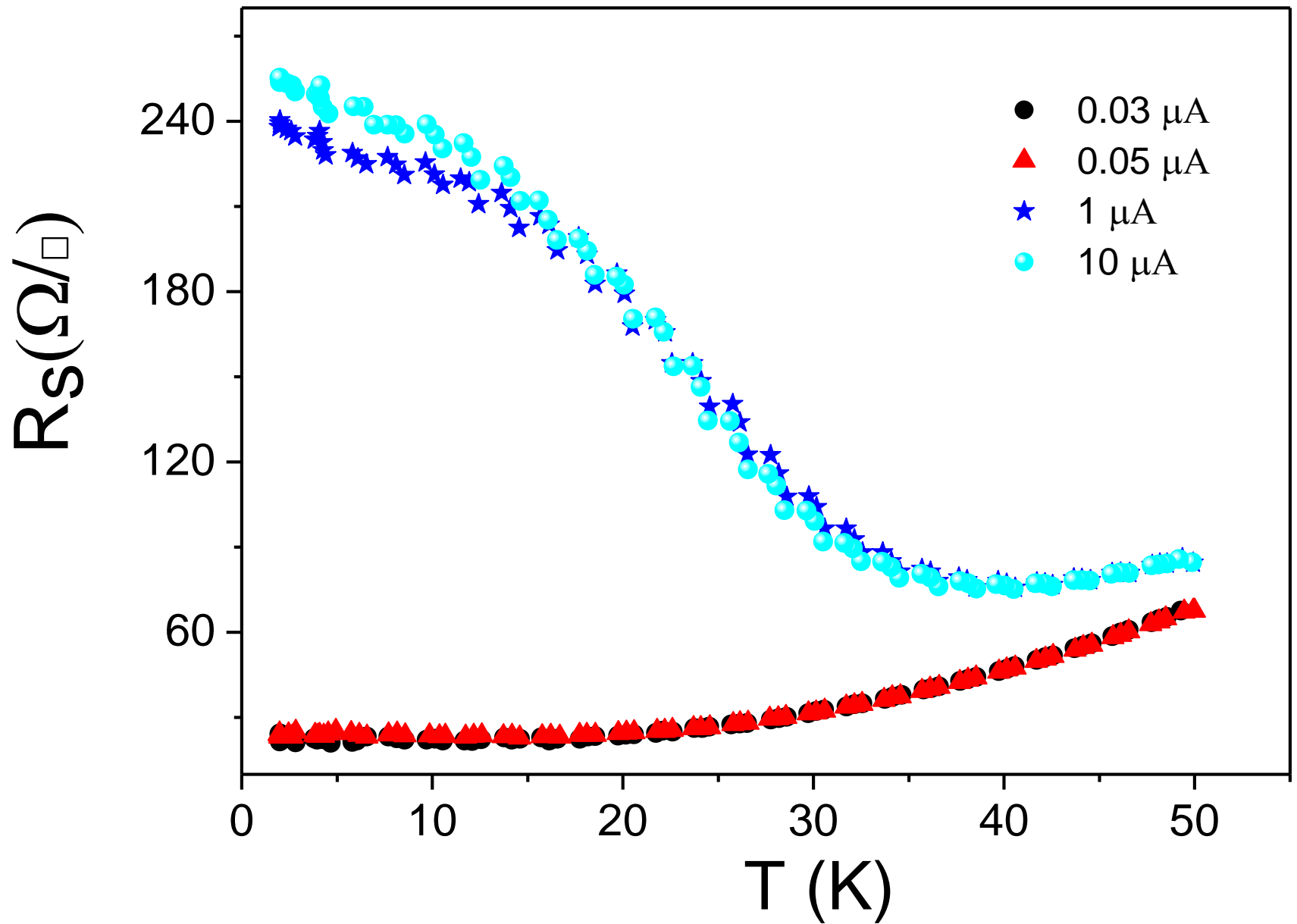

FIG. 3: Temperature dependence of $R_s$ during cooling from 50K to 5K, with different applied currents.

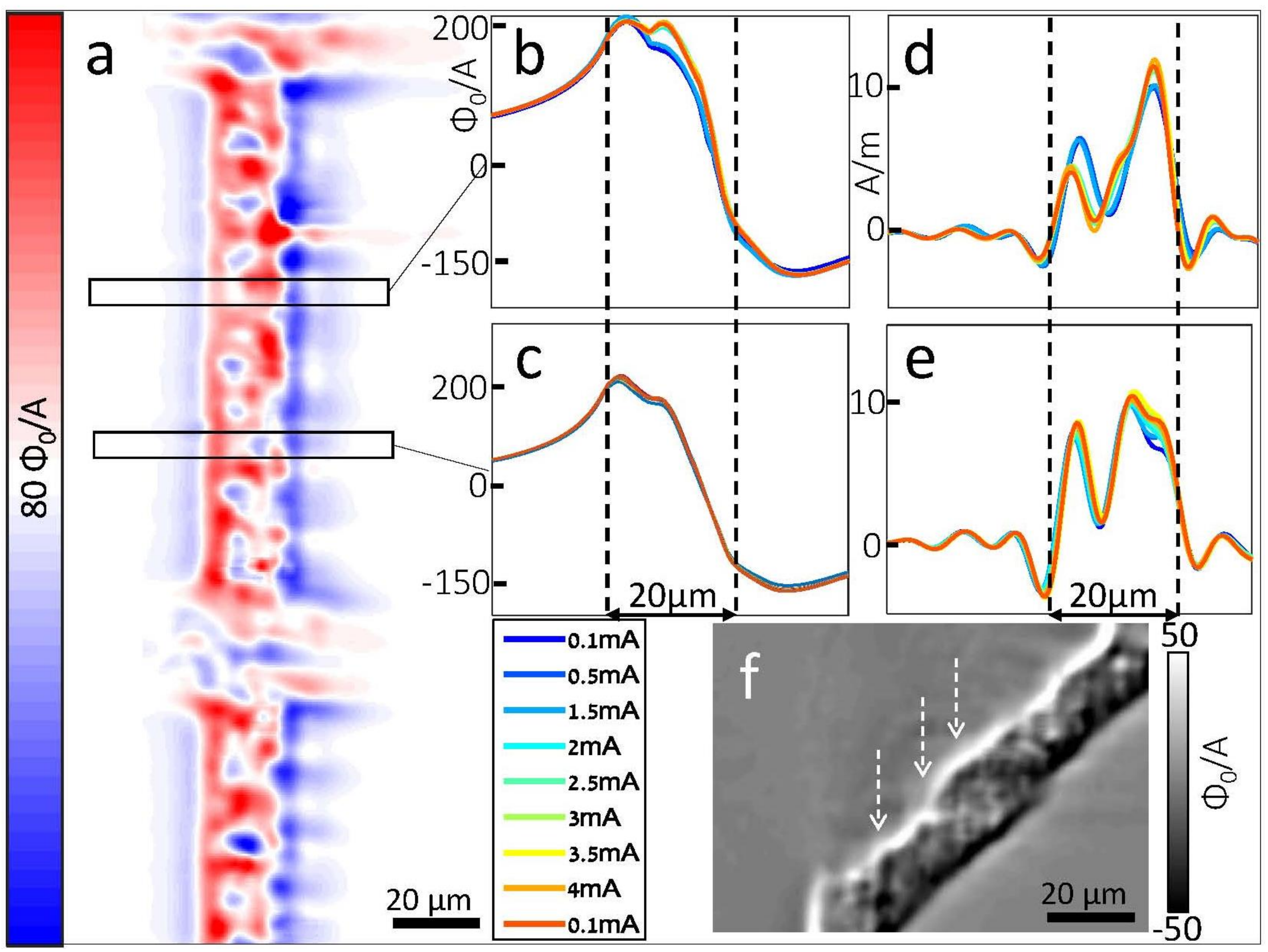

FIG4: (a) Initial image of 20 μm bridge. (b)-(c) line cuts as a function of current in the marked region in Figure 4(a). (d)-(e) Two-dimensional current reconstruction of Figure 4(b) and (c) respectively.(f) High resolution SQUID image of a particular region showing domain structure masked by the inhomogeneous current flow. The arrows are the guide to the eyes for identifying the masked domains.

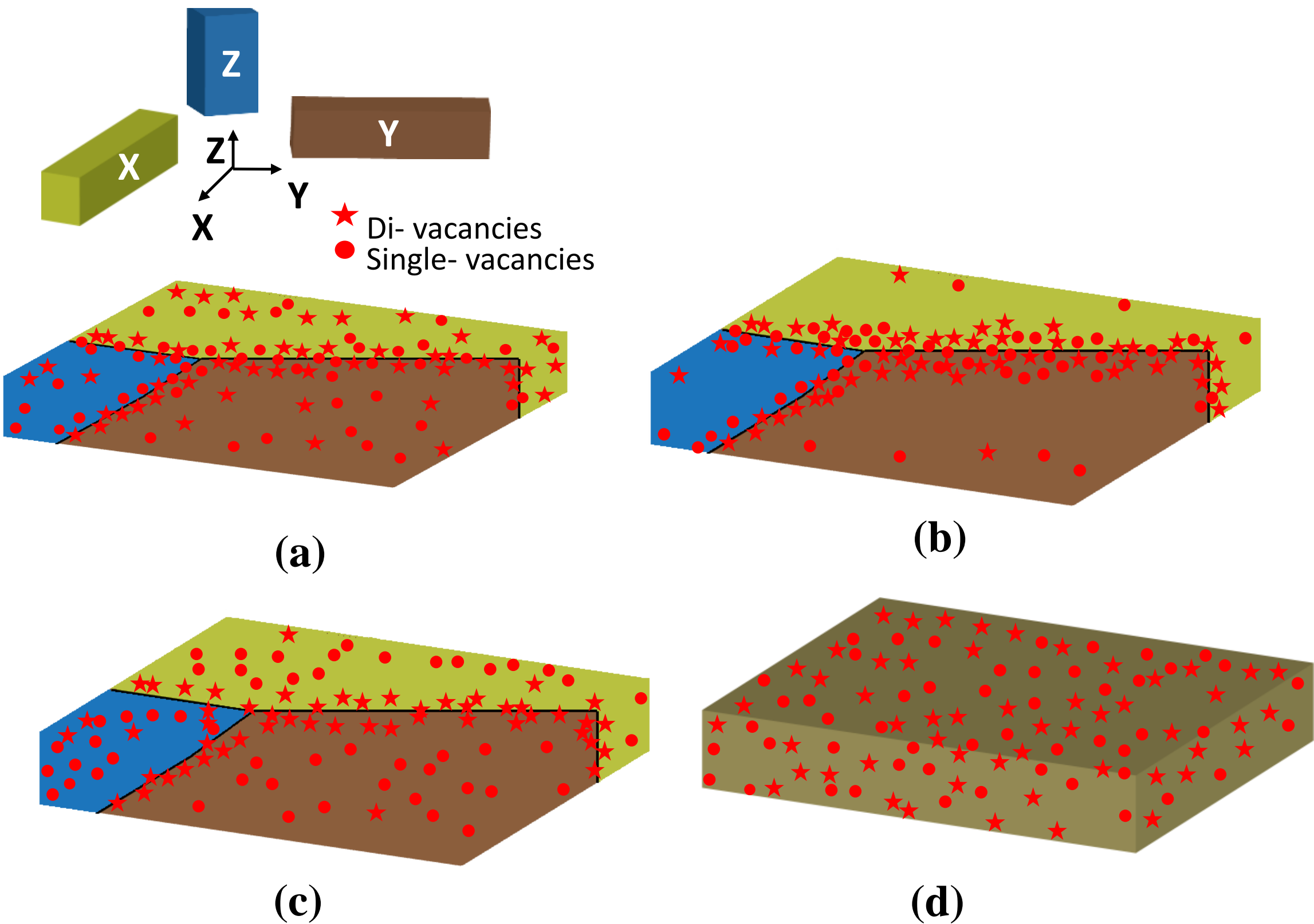

FIG 5: (a) Schematics of tetragonal domains in $SrTiO_3$ and the formation of twin boundaries at the intersection of three possible domains with different orientation. (The domains are labelled as X, Y and Z according to their long axis). During initial cool down from room temperature, oxygen vacancies of two types, single and di-vacancies become static below 70 K (single vacancies) and 160 K (di-vacancies). The vacancies have tendency to be close to the DWs and they have a small effect on the conductivity of the 2DEG. (b) The electrical fields associated with currents above a certain threshold induce below 40 K a ferroelectric transition of the DWs. The polarization of the DWs increases the density of oxygen vacancies in the vicinity of the DWs. The associated electrical fields in the vicinity of the DWs act as internal gating which yields a noticeable decrease in the conductivity the 2DEG. Above 40 K and below 70 K the DWs are not polarized; however, the vacancies are still static and therefore the initially higher conductivity is not recovered. (c) Upon warming, above ~70 K the single oxygen vacancies become mobile which yields a partial conductivity recovery. (d) Upon further warming, above~ 160 K the oxygen di-vacancies become mobile which yields a full conductivity recovery. Notably at this temperature the tetragonal domain structure is not evident as tetragonal to cubic phase transition occurs in $SrTiO_3$ at ~105K.